\documentclass[pra,superscriptaddress,showpacs,twocolumn,10pt]{revtex4}
\usepackage{amssymb}
\usepackage{graphicx}

\def\>{\rangle}

\begin{document}

\title{Quantum teleportation using cluster states}
\author{B. Zeng}
\affiliation{Department of Physics, Tsinghua University, Beijing, 100084, China}
\author{D.L. Zhou}
\affiliation{Center for Advanced Study, Tsinghua University, Beijing 100084, China}
\author{Z. Xu}
\affiliation{Department of Physics, Tsinghua University, Beijing, 100084, China}
\author{C.P. Sun}
\email{suncp@itp.ac.cn}
\homepage{http:// www. itp.ac.cn/~suncp}
\affiliation{Institute of Theoretical Physics, The Chinese Academy of Sciences, Beijing,
100080, China}
\date{\today}

\begin{abstract}
A protocol of quantum communication is proposed in terms of the
multi-qubit quantum teleportation through cluster states (Phys.
Rev. Lett. \textbf{86}, 910 (2001)). Extending the cluster state
based  quantum teleportation  on the basic unit of three qubits
(or qudits), the corresponding multi-qubit network is constructed
for both the qubits and qudits (multi-level) cases. The classical
information costs to complete this communication task is also
analyzed. It is also shown that this quantum communication
protocol can be implemented in the spin-spin system on lattices.
\end{abstract}

\pacs{3.67.Lx, 3.67.-a}
\maketitle

Quantum communication, a way of communication based on various features of
quantum coherence (such as quantum entanglement and quantum non-cloning,
which undoubtedly have no counterparts in classical worlds), are more
reliable and efficient than any classical communication method\cite{Ni}.
Quantum teleportation \cite{Be} is one of the most remarkable protocols
among those tasks of quantum communication, which completes the striking
task of remotely preparing an unknown quantum state of a particle without
sending this particle itself. Since proposed by Bennett et. al, quantum
teleportation has attracted much attention from both experimentalists \cite%
{Bou,Kimble,Mati,Kim,Peng,Dur} and theoreticians \cite{Dur,Wer,Yu,Zeng}.
Appropriate choice and measurements of entangled states are crucial to the
implementation of those quantum communication protocols.

Recently, a novel kind of entangled states - cluster states was introduced
with their remarkable properties such as maximal connectedness and
persistency of entanglement\cite{Bri}. They can be used to build a one-way
universal quantum computer associated with only single-qubit measuments\cite%
{Rau}. It was also pointed out that those computers can be implemented
physically since the creation of the cluster states needs only the
Ising-type interations\cite{Bri}, which can be easily found in various
condensed matter systems with proper spin-spin coupling on lattices\cite{Bri}
and even in the cold atom system in optical lattices\cite{Du}. Very
recently, with an intrinsically algebraic consideration, we generalized the
concept of cluster states to the more general case with qudits (multi-level
systems) and then demonstrate their quantum correlation features and the
fundamental computational properties in many aspects \cite{Zzxs}.

In this letter the cluster state is chosen as the source of
entanglement to propose a novel protocol of quantum communication.
The essence and functioning of this quantum teleportation scheme
using the cluster state can be viewed as an identity quantum logic
gate, which was originally used in the schem of measurement-based
one-way quantum computers\cite{Bri} only based on the cluster
state and the corresponding single quantum measurements. The
advantage of this implementation is its scalability: a basic
\textquotedblleft unit" can be easily constructed \ for a smallest
(three-qubit) quantum logic network and then this unit can be
naturally enlarged to form a more-qubit network. Moreover, this
protocol can be realized in the spin-spin system on lattices,
which physically guarantees the scalability and implies that the
communication can be carried out from certain lattice sites to
those at a long distance. With those considerations
we will explicitly build the quantum network for a quantum teleportation of $%
N$-qubits and $N$-qudits. We also analyze the entanglement and classical
information costs to complete the teleportation task on a cluster.

We first view that the original \ procedure of quantum
teleportation protocol mainly depends on quantum measurement with
respect to four Bell states, the complete set of maximally
entangled state of two particles. It was discovered that the Bell
states can be produced by a proper interaction from certain
factorized states. In fact, using the Pauli matrices $\sigma
_{s}(s=x,y,z)$ defined for qubit states $|0\rangle $ and
$|1\rangle $, we write the tensor product operator\cite{Rau1}
$S^{ab}=\frac{1}{2}(I+\sigma _{z}^{a}+\sigma _{z}^{b}-\sigma
_{z}^{a}\otimes \sigma _{z}^{b})$ acting on particles $a$ and $b$.
Experimentally, this can be easily realized as a
time evolution governed by the Ising-type interaction Hamiltonian, i.e. $%
H_{I}=\hbar g\sum\limits_{b-a=1}\sigma _{z}^{a}\sigma _{z}^{b}$ for a
spin-lattice systems\cite{Bri} or a cold atom system in optical lattices\cite%
{Du}. Define $|\pm \rangle _{s}=\frac{1}{\sqrt{2}}(|0\rangle _{s}\pm
|1\rangle _{s})$($s=1,2,3,..)$. It is easy to check that $S^{ab}$ transfer
four factorized states $\{|\pm \rangle _{a}\otimes |\pm \rangle _{b}$,$|\pm
\rangle _{a}\otimes |\mp \rangle _{b}\}$ of particles $a$ and $b$ into four
Bell basis vectors $|B_{\pm 1}(ab)\rangle =\frac{1}{\sqrt{2}}(|0\rangle
_{a}|-\rangle _{b}\pm |1\rangle _{a}|+\rangle _{b})$ and $|B_{\pm
0}(ab)\rangle =\frac{1}{\sqrt{2}}(|0\rangle _{a}|+\rangle _{b}$ $\pm $ $%
|1\rangle _{a}|-\rangle _{b})$. Therefore, for a given two
particle wave function $|\Phi \rangle _{ab}$, the quantum
measurements on its unitary
transformation $S^{ab}$ $|\Phi \rangle _{ab}$ with respect to four states $%
|\pm \rangle _{a}\otimes |\pm \rangle _{b}$ and $|\pm \rangle
_{a}\otimes |\mp \rangle _{b}$ are exactly equivalent to a
Bell-measurement with respect to $|B_{\pm 1}(ab)\rangle $ and
$|B_{\pm 0}(ab)\rangle $. In this sense the results of the single
particle measurements on $|\pm \rangle _{a}$ and $|\pm \rangle
_{b}$ will determinate the Bell-measurements on $|\Phi \rangle
_{ab}$. We call this \textquotedblleft two-step Bell-measurement".

With the above observations we now turn to consider how to
teleport a state of a single particle located in site $1$ of the
lattice to another particle located in site $3$. Since the
controlled evolution governed by the above mentioned
\textquotedblleft Ising" type interaction plus the single particle
measurements can realize a two-step Bell-measurement, a quantum
telportation can be naturally viewed as an identity quantum logic
gate on the one way quantum compute. To be more concrete, we
consider the simplest quantum logic network which can implement
the identity gate\cite{Rau1} shown in Fig. 1. There is a linear
array of three sites on lattice and three spins attached on these
sites interact with each other through the \textquotedblleft
Ising" type interaction. For this physical system we can propose
our protocol of quantum teleportation using cluster states in
three steps:
\begin{figure}[tbph]
\begin{center}
\includegraphics{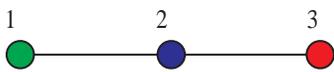}
\end{center}
\caption{The cluster for quantum teleportation of one qubit state.
The circle with the number $n$ denotes the $n$-th qubit, and the
two qubits connected by a line are neighbours. The green, blue and
red colors are used to mark the input, body and output parts of
the cluster respectively. }
\end{figure}

I. Initially, Alice holds two particles $1$ and $2$ and let the particle $1$
be in an unknown state $|\psi \rangle _{in}=|\pi (\alpha ,\beta )\rangle =$ $%
\alpha |0\rangle _{1}+\beta |1\rangle _{1}$ (where $|\alpha |^{2}+|\beta
|^{2}=1$) and the particle $2$ in the state $|+\rangle _{2}$. Bob holds
particle $3$ in the state $|+\rangle _{3}$. After an evolution derived by
the Ising type interaction for a length time $t$ (governed by the equation $%
\pi =\int g(t)dt$), the unitary transformation $S=S^{12}S^{23}$ on the
initial product state$|\Psi (0)\rangle =|\pi _{1}(\alpha ,\beta )\rangle
\otimes |+\rangle _{2}\otimes |+\rangle _{3}$ will result in a three
particle entangled state $|\Psi (t)\rangle $. It is noticed that, since $%
S^{12}$ commutes with $S^{23}$, the unitary transformation $S$ can be
decomposed into two independent sub-steps.

II. Alice measure $\sigma _{x}^{1}$ and $\sigma _{x}^{2}$ for her particles $%
1$ and $2$ in the single-particle basis $|\pm \rangle $ respectively and
then tell Bob which $(x_{1,}x_{2})$ of the four possible results about $%
\frac{1}{2}(\sigma _{x}^{(1)}+1$ $)$ and $\frac{1}{2}(\sigma
_{x}^{(2)}+1$ $)$, where $x_{1},x_{2}\in {(0,1)\sim }({+,-)}$. In
this sense the total wave function $|\Psi (t)\rangle $ collapses
onto
\begin{equation}
|\Psi _{3}(t)\rangle \propto \langle x_{1,}x_{2}|S^{12}S^{23}|\Psi
(0)\rangle =\langle B_{\eta }|S^{23}|\Psi (0)\rangle .
\end{equation}%
That is just the projection of $|\Phi (t)\rangle =S^{23}|\Psi (0)\rangle $
onto the Bell basis $|B_{\eta }\rangle =S^{12}|x_{1,}x_{2}\rangle $, where
the role of $S^{23}$ (which transforms the factorized state $|+\rangle
_{2}\otimes |+\rangle _{3}$ to a maximally entangled state $\frac{1}{2}%
(|0\rangle _{2}|-\rangle _{3}+|1\rangle _{2}|+\rangle _{3})$) is to
distribute an entanglement source to Alice and Bob.

III. According to this result $(x_{1},x_{2})$ Bob make an $(\alpha ,\beta
)-independent$ unitary transformation
\begin{equation}
U_{\Sigma }^{\dag }=(\sigma _{x}^{(3)})^{x_{2}+1}(\sigma
_{z}^{(3)})^{x_{1}+1}
\end{equation}
on his particle $3$ to obtain unknown state $|\pi _{3}(\alpha
,\beta )\rangle =\alpha |0\rangle _{3}+\beta |1\rangle _{3}$ for
particle $3$ exactly. In fact, according to the \textquotedblleft
one way quantum computer " theory\cite{Rau1}, we have the
measurement induced output $|\psi \rangle _{out}=U_{\Sigma }|\psi
\rangle _{in}$. Here, $U_{\Sigma }$ is the extra rotation we
should carry out in order to finish our task of teleportation.
Since we measure particles $1$ and $2$ in $\sigma _{x}$ basis,
$U_{\Sigma }$ must belong to the Pauli algebra generated by
$\sigma _{s}^{(3)}(s=1,2,3)$ and thus has the form $U_{\Sigma
}=(\sigma _{z}^{(3)})^{x_{1}+1}(\sigma _{x}^{(3)})^{x_{2}+1}$,
which is determined by the outcomes of measurement results ($+,-$)
in $\sigma _{x}$ basis. Thus a one-way unitary transformation
$U_{\Sigma }^{\dag }$ is sufficient to transform particle $3$
to$|\pi _{3}(\alpha ,\beta )\rangle \sim |\psi \rangle _{in}$.

It is noticed that the steps I and II together implement an
identity quantum logic gate on the cluster. However, it should be
emphasized that the quantum teleportation cannot be implemented
only using cluster state and one-qubit measurement. Indeed, the
step III must be carried out, i.e. we should make a unitary
transformation on the output particle $3$ to ensure that the state
of particle $3$ is exactly that one of particle $1$. This is the
main difference between quantum teleportation and quantum
computation based on cluster state. In fact, in the
measurement-based quantum computation, we only need the output
$|\psi \rangle _{out}=U_{\Sigma }|\psi \rangle _{in}$ (other than
$|\psi \rangle _{in}$) itself to readout the result of computing.
Finally we come to the conclusion that, to implement quantum
teleportation on a cluster is equivalent to realize an identity
quantum logic gate on the same cluster. Thus, our task remains to
construct a quantum logic network to implement an identity quantum
logic gate and find the explicit form of the unitary
transformation $U_{\Sigma }$ in terms of the outcomes of the
measurements.

Based on the detailed analysis above, it is straightforward to suggest a
general protocol for cluster-based quantum teleportation to transport $multi$%
-qubit information between the two sets of qubits located on the
lattice sites with a long distance. Essentially, it is a general
identity quantum logic gate realized with a quantum network formed
by a cluster plus the appropriate single particle measurements (as
Fig.2). This cluster-based quantum network can be formally divided
into three parts, the input $(I)$ unit with $N$ qubits, the output
unit $(O)$ with $N$ qubits and the connection block $(B)$-the body
part. When we input a $N$ qubit state $|\psi \rangle _{in}$, the
output $|\psi \rangle _{out}=U_{\Sigma }|\psi \rangle _{in}$ is
induced by the single qubit measurements performed simultaneously
on all sites of $I$ and $B$. This seems simple, but one should pay
attention to the corresponding problem of information cost. As
well-known, to teleport an arbitrary $N$-qubits states, at least
$N$ ebits ($N$ entanglement pairs) and $2N$ classical bits of
information must be used. Can we reach this tightly lower bound
using an entangled cluster state? The answer is affirmative and we
will also show how to build the quantum logic network to realize
such quantum communication with the tightly lower bound of cost of
information as follows. A one-dimensional chain cluster is
sufficient to realize our entanglement sources. In fact, we can
use ordered indices to label qubits in the one-dimensional cluster
of $M($even number$)$ qubits. Therein the qubits with odd and even
indices (odd- and even-bit)can form two subsystems with $M/2$
qubits respectively. Then a maximally entangled state can form
between these two subsystems. This conclusion can be reached by
observing that, by measuring all the qubits with odd indices in
the chain cluster with respect to the $\sigma _{z}$ -basis, all
the qubits with even indices will be projected onto\ the $\sigma
_{x}$ -basis. Then there is a one-to-one correspondence between
the measurement result and the projection result. Thus, we will
have $N$-ebits of entanglement if we use a one-dimensional cluster
state of $2N$-qubits and \ give the odd-qubits to Alice and the
even-qubits to Bob.

\begin{figure}[tbph]
\begin{center}
\includegraphics[height=8cm]{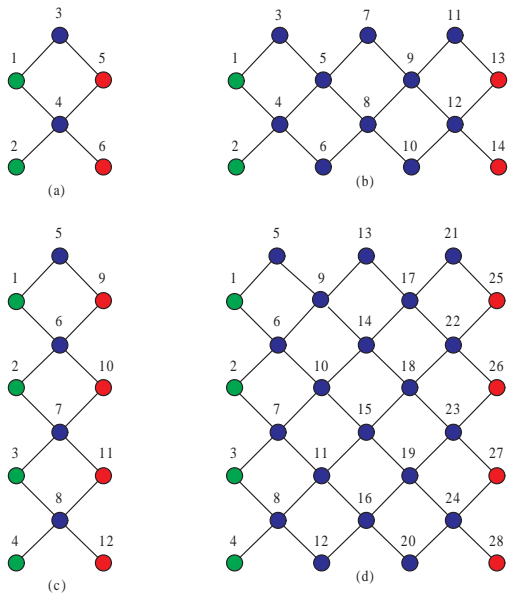}
\end{center}
\caption{The clusters for quantum teleportation of qubit states.
The circle with the number $n$ denotes the $n$-th qubit, and the
two qubits connected by a line are neighbors. The green, blue and
red colors are used to mark the input, body and output parts of
the cluster respectively.(a)The cluster to quantum teleport the
state of qubits $1$ and $2$ to that of $3$ and $4$; (b)The
horizontal expansion of the cluster in (a);(c)The vertical
expansion of the cluster in (a);(d)The lattice for general quantum
teleportation.} \label{fig2}
\end{figure}

With above general consideration for the information cost in the
cluster-based teleportation of $N$-qubits, we can embark on constructing a
proper cluster to implement quantum teleportation of unknown $N$ qubit
state. Without loss of generality, we consider the quantum teleportation of $%
2 $-qubit state on the six qubit cluster as shown in Fig. \ref{fig2} (a).

According to Ref. \cite{Rau1}, the corresponding cluster state is
defined by the following equations:
\begin{eqnarray}
\sigma _{x}^{(1)}\sigma _{z}^{(3)}\sigma _{z}^{(4)}|\phi \rangle &=&|\phi
\rangle ,  \label{x1} \\
\sigma _{x}^{(2)}\sigma _{z}^{(4)}|\phi \rangle &=&|\phi \rangle ,
\label{x2} \\
\sigma _{z}^{(1)}\sigma _{x}^{(3)}\sigma _{z}^{(5)}|\phi \rangle &=&|\phi
\rangle ,  \label{x3} \\
\sigma _{z}^{(1)}\sigma _{z}^{(2)}\sigma _{x}^{(4)}\sigma _{z}^{(5)}\sigma
_{z}^{(6)}|\phi \rangle &=&|\phi \rangle ,  \label{x4} \\
\sigma _{z}^{(3)}\sigma _{z}^{(4)}\sigma _{x}^{(5)}|\phi \rangle &=&|\phi
\rangle ,  \label{x5} \\
\sigma _{z}^{(4)}\sigma _{x}^{(6)}|\phi \rangle &=&|\phi \rangle .
\label{x6}
\end{eqnarray}%
The above system of equation can be simplified as
\begin{eqnarray}
\sigma _{x}^{(1)}\sigma _{x}^{(5)}|\phi \rangle &=&|\phi \rangle ,
\label{x15} \\
\sigma _{x}^{(2)}\sigma _{x}^{(6)}\sigma _{z}^{(6)}|\phi \rangle &=&|\phi
\rangle ,  \label{x26} \\
\sigma _{z}^{(1)}\sigma _{x}^{(3)}\sigma _{z}^{(5)}|\phi \rangle &=&|\phi
\rangle ,  \label{z15} \\
\sigma _{z}^{(2)}\sigma _{x}^{(3)}\sigma _{x}^{(4)}\sigma _{z}^{(6)}|\phi
\rangle &=&|\phi \rangle .  \label{z26}
\end{eqnarray}%
For the measurement pattern \cite{Rau1}$\{\sigma _{x}^{(1)},\sigma
_{x}^{(2)},\sigma _{x}^{(3)},\sigma _{x}^{(4)}\}$, we obtain the measurement
result $\{s_{1},s_{2},s_{3},s_{4}\}$. Based on the theorem $1$ in Ref. \cite%
{Rau1}, we conclude that the realized unitary transformation in the two
particle Hilbert space of qubit $5$ and $6$ is an identity $I$ . \
Correspondingly the extra rotation $U_{\Sigma }$ is
\begin{equation}
U_{\Sigma }=(\sigma _{z}^{(5)})^{s_{1}}(\sigma _{x}^{(5)})^{s_{3}}(\sigma
_{z}^{(6)})^{s_{2}}(\sigma _{x}^{(6)})^{s_{3}+s_{4}}.
\end{equation}%
After obtaining the measurement results through classical
channels, one needs to perform the unitary gate $U_{\Sigma }^{\dag
}$ in the output unit. Then the input state is determinately
teleported to the output unit.

With the aid of the connecting natures of the cluster state, the
above quantum teleportaion protocol for $2$-qubit state can be
generalized directly to the case with $N$-qubit state. Along the
vertical direction, we extend the above cluster to that with $3N$
qubits, shown as in Fig. \ref{fig2} (c). After measuring Pauli
operator $\sigma _{x}$ in the input unit and the connection block
of the cluster, thereafter performing the proper unitary
transformation, depending on the former measurement results, the
quantum teleportaion of $N$- qubit state is completed from the
input unit to output part. Another way to extend the protocol is
to quantum teleport a state to a longer distance through many
medial qubits. Contrast to the above extension along the vertical
direction, we extend the cluster along the horizontal direction as
in Fig. \ref{fig2} (b). This expansion might be understood as
connection of the basic units of quantum teleportation in
sequence. Thus, to complete such quantum teleportation, we also
need to measure $\sigma _{x}$ for all the qubits in the input unit
and the body part of the cluster, and then to perform a
measurement-dependent unitary transformation in the output part.
It must be emphasized that {\emph{only the collective measurement
results (more precisely, the sums of the measured values of
$\sigma _{x}$ in every horizontal line)}} can influence the final
unitary transformation in the output part. Accordingly, the
classical information cost to implement this teleportation
protocol for every one qubit state is two bits, which is
exactly the same as that in the original protocol suggested by Bennett et al%
\cite{Be}.

Extending the cluster in both vertical and horizontal directions, we obtain
a two dimensional lattice array of qubits as demonstrated in Fig. \ref{fig2}
(d). Obviously, the same procedure as above can complete quantum
teleportaion on this lattice. In addition, if provided with a large lattice
of spins, we can chose the proper cluster by measuring $\sigma _{z}$ to
delete the redundant qubits from the lattice. Then, as long as the cluster
for quantum teleportation can be formed in the lattice, we can teleport the
state of the input qubits to the output qubits, whichever sites they locate.

So far our investigation has centered in how to teleport the state
of qubits using cluster state. Then a natural question arises: Can
we quantum teleport the state of qudits ($d$-level system) in a
similar way? To teleport the state of qudits, it is indispensable
to generalize the concept of cluster state of qubits \ for qudits.
Most recently, we have just carried out this generalization
successfully \cite{Zzxs} and proved two useful theorems parallel
to that in Ref. \cite{Rau1}. With this generalization we can
parallel propose the protocol of qudit based quantum
teleportation.

Let us begin with the simplest case---the quantum teleportation of one qudit
state. We take a cluster of three qudits as Fig.1. The higher dimensional
cluster state \cite{Zzxs} is defined by
\begin{eqnarray}
X_{1}^{\dagger }Z_{2}|\phi \rangle _{\mathcal{C}} &=&|\phi \rangle _{%
\mathcal{C}},  \label{3d1} \\
Z_{1}X_{2}^{\dagger }Z_{3}|\phi \rangle _{\mathcal{C}} &=&|\phi \rangle _{%
\mathcal{C}},  \label{3d2} \\
Z_{2}X_{3}^{\dagger }|\phi \rangle _{\mathcal{C}} &=&|\phi \rangle _{%
\mathcal{C}},  \label{3d3}
\end{eqnarray}%
where the generalized Pauli operators$Z$ and $X$ are the generators of
quantum plane algebra with $q^{d}=1$ \cite{Sun}. The $Z$-diagonal
representation of $Z$ and $X$ given by
\begin{eqnarray}
Z &\equiv &\sum_{k_{0}}^{d-1}|k\rangle q_{d}^{k}\langle k|, \\
X &\equiv &\sum_{k=0}^{d-1}|k\rangle \langle k+1|,
\end{eqnarray}%
for $q_{d}=e^{i\frac{2\pi }{d}}.$From Eqs. \ref{3d1}-\ref{3d3}, we obtain
\begin{eqnarray}
X_{1}X_{3}^{\dagger }|\phi \rangle _{\mathcal{C}} &=&|\phi \rangle _{%
\mathcal{C}}, \\
Z_{1}^{\dagger }X_{2}Z_{3}^{\dagger }|\phi \rangle _{\mathcal{C}} &=&|\phi
\rangle _{\mathcal{C}}.
\end{eqnarray}

When we obtain the measurement results $\{s_{1},s_{2}\}$ for the measurement
pattern $\{X_{1},X_{2}\}$, the fundamental theorem $2$ \ proved in Ref. \cite%
{Zzxs} determines the unitary transformation on the third qudit $UU_{\Sigma }
$. Here, $U$ is defined by
\begin{equation}
UX_{3}U^{\dagger }=X_{3}^{\dagger },UZ_{3}U^{\dagger }=Z_{3}^{\dagger },
\end{equation}%
and $U_{\Sigma }=Z_{3}^{-s_{1}}X_{3}^{s_{2}}$. To complete this quantum
teleportation, one needs to perform a unitary transformation $U_{\Sigma
}^{\dagger }U$ on the third qudit. Therefore, in the same way as in the
qubit case, the above one-qudit quantum teleportation scheme can be extended
to the quantum teleportation of $N$-qudit state , and even be extended to
the quantum teleportation of qudit states through a large two dimensional
lattice.

{\vskip2mm} To sum up, we have universally proposed a novel protocol for
quantum teleportation of qubits and qudits based on cluster states.
According to the analysis about the properties of entanglement resources, a
cluster composed of $3N$ qubits (or qudits) suffices to teleport a $N$ qubit
(or qudit) state. When the cluster is extended to form a two or higher
dimensional lattice, this protocol of quantum teleportation can still work
well.

{\vskip 2mm}\textbf{Acknowledgements}

The work of D. L. Z is partially supported by the National Science
Foundation of China (CNSF) grant No. 10205022. The work of Z.X is also
supported by CNSF (Grant No. 90103004, 10247002).The work of C.P.S is
supported by the CNSF( grant No.90203018) and the knowledged Innovation
Program (KIP) of the Chinese Academy of Science and the National Fundamental
Research Program of China with No 001GB309310.


\begin{thebibliography}{99}
\bibitem{Ni} M. A. Nielsen and I. S. Chuang, \textit{Quantm computation and
quantum information}, Cambridge University Press (2000).

\bibitem{Be} C. H. Bennett, G. Brassard, C. Crepeau, R. Jozsa, A. Peres, and
W. K. Wooters, Phys. Rev. Lett.70, 1895 (1993).

\bibitem{Bou} D. Bouwmeester, et al., Nature \textbf{390}, 575 (1997).

\bibitem{Kimble} A. Furusawa, J. L. Sorensen, S. L. Braunstein, et al.
Science \textbf{282} 706, (1998).

\bibitem{Mati} D. Boschi, S. Branca, F. De Martini, L. Hardy, S. Popescu,
Phys. Rev. Lett. \textbf{80} 1121, (1998).

\bibitem{Kim} Y. H. Kim, S. P. Kulik, and Y. H. Shi, Phys. Rev. Lett.
\textbf{86}, 1370 (2001).

\bibitem{Peng} J. Zhang, C. D. Xie, K. C. Peng, Phys. Lett. \textbf{A287}, 7
(2001).

\bibitem{Dur} W. D\"{u}r and J. I. Cirac, J. Mon. Opt. \textbf{47}, 247
(2000).

\bibitem{Wer} R. F. Werner, quant-ph/0003070 (2000).

\bibitem{Yu} S. X. Yu and C. P. Sun, Phys. Rev. \textbf{A61}, 2310(2000).

\bibitem{Zeng} B. Zeng, X. S. Liu, Y. S. Li and G. L. Long, Commun. Theor.
Phys. \textbf{38}, 537 (2002).

\bibitem{Bri} H. J. Briegel and R. Raussendorf, Phys. Rev. Lett. \textbf{86}%
, 910 (2001).

\bibitem{Rau} R. Raussendorf and H. J. Briegel, Phys. Rev. Lett. \textbf{86}%
, 5188 (2001).

\bibitem{Du} L. M. Duan, E. Demler, and M. D. Lukin, cond-mat/0201564 (2002).

\bibitem{Rau1} R. Raussendorf, D. E. Browne, and H. J. Briegel,
quant-ph/0301052 (2003).

\bibitem{Zzxs} D. L. Zhou, B. Zeng, Z. Xu, C. P. Sun, quant-ph/0304054.

\bibitem{Sun} C. P. Sun, in ``Quantum Group and Quantum Integrable Systems",
ed by M. L. Ge, World Scientific, 1992, p.133; M. L. Ge, X. F. Liu, C. P.
Sun, J. Phys A-Math. Gen \textbf{25} (10): 2907, (1992).
\end{thebibliography}
\end{document}